# Towards Commercializing Vanadium Dioxide Films: Investigation of the Impact of Different Interface on the Deterioration Process for Largely Extended Service Life

Tianci Chang,[1,2] Xun Cao,[1,*] Ning Li,[3] Shiwei Long,[1,2] Ying Zhu,[1,2] Jian Huang,[1] Hongjie Luo,[4] and Ping Jin[1,5,**]

[1]State Key Laboratory of High Performance Ceramics and Superfine Microstructure, Shanghai Institute of Ceramics, Chinese Academy of Sciences, Shanghai 200050, China

[2]University of Chinese Academy of Sciences, Beijing 100049, China

[3]Department of Materials Science and Engineering, College of Science, China University of Petroleum Beijing, Beijing 102249, China

[4]School of Materials Science and Engineering, Shanghai University, Shanghai 200444, China

[5]National Institute of Advanced Industrial Science and Technology (AIST), Moriyama, Nagoya 463-8560, Japan

*Correspondence: cxun@mail.sic.ac.cn

**Correspondence: p-jin@mail.sic.ac.cn

**SUMMARY**

**Long-term stability is the most pressing issue that impedes commercialization of Vanadium Dioxide ($VO_2$) based functional films, which show a gradual loss of relative phase transition performance, especially in humid conditions when serving as smart windows. Here, we investigated the impact of different interface on the deterioration process of $VO_2$ films and proposed a novel encapsulation structure for largely extended service life. Hydrophobic and stable hafnium dioxide ($HfO_2$) layers have been incorporated with $VO_2$ films for encapsulated surfaces and cross-sections. With modified thickness and structure of $HfO_2$ layers, the degradation process of $VO_2$ can be effectively suppressed. The proposed films can retain stable phase transition performances under high relative humidity (90%) and temperature (60℃) over 100 days, which is equal to ~16 years in the real environment. Improving the stability of $VO_2$ materials is a necessary step towards commercializing production of high-performance films for long-term use.**

**Context & Scale**

With the shortage of existing energy and the demand for new energy, vanadium dioxide ($VO_2$) has sparked numerous potential applications as sensors and switches for optoelectronic devices and thermochromic coatings for energy-efficient effect, due to the significant changes in electoral and optical properties during the semiconducting-metallic transition (SMT) process. Unfortunately, $VO_2$ would gradually transform into other compounds, which does not possess the phase transition performance near room temperature. The vast majority of research on vanadium dioxide over the past four decades has focused on its phase transition properties and relative electrical and optical performances. However, previous work hardly concentrates on the stability of $VO_2$, which is the main challenge of commercializing application. In this work, we investigated the impact of different interface on the deterioration process of $VO_2$ films and proposed a novel encapsulation structure for largely extended service life.

## INTRODUCTION

As an archetypal phase transition material, vanadium dioxide ($VO_2$) has been widely investigated due to the significant change in physical properties at a critical transition temperature ($T_c$, ~68°C) from monoclinic phase $VO_2$ (M) to rutile phase $VO_2$ (R).[1] Accompanied with the transition process, $VO_2$ shows a dramatic modification of the conductivity on several orders of magnitude[2] and the optical properties from infrared transparent to infrared opaque[3] with different external stimuli, including temperature, light, strain, electric-field and terahertz-field.[2-6] Due to the unique phase transition, $VO_2$ has sparked numerous potential applications as field-effect transistors,[7] sensors,[8-10] metamaterials[11-13] and thermochromic coatings.[14-16] Particularly, $VO_2$ based thermochromic coatings have been widely investigated due to their facile fabrication methods, simple structures, and automatic control of solar modulation for energy-efficient effect depending on environmental temperatures without additional energy stimulus.[17-19]

When talking about commercialization of $VO_2$ based thermochromic coatings, the main parameters are the transition temperature, optical performances (including solar modulation ability (($\Delta T_{sol}$) and luminous transmittance ($T_{lum}$)), production costs and long-term stability (service life).[20-23] In recently reported work, researchers have made many efforts to solve above obstacles. The transition temperatures can be successfully decreased by doping and strain modulation.[24-26] By fabricating multilayer structures, optical performances can be enhanced with excellent solar modulation abilities larger than 15% accompanying by acceptable luminous transmittances for effective energy-efficient effect.[18, 27, 28] To reduce the production costs, low-temperature deposition technologies have been developed, such as post-annealing and using buffer layers.[29-31]

More importantly, a certain lifetime of $VO_2$-based films is required of final products for commercializing application, and the prerequisite for this is the demonstration of certain stability of the materials employed.[18, 22] Unfortunately, $VO_2$ is considered to be unstable in the real environment, especially in humid conditions with moisture. $VO_2$ will gradually transform into the most thermodynamically stable phase of vanadium oxide ($V_2O_5$) or their hydroxides and lead to a quick decrease of its phase transition properties.[32] A common way to resolve this problem is utilizing protective to keep $VO_2$ layers from oxidation and deterioration due to the reaction with the moisture in the air.[32-34] Nevertheless, with respect to these work, there is a lack of systematic illustration of the deterioration process of $VO_2$ films and the reported work cannot achieve desirable stability for long time use. Meanwhile, additional protective layers always lead to deteriorated optical performance of $VO_2$ layers with decreased solar modulation abilities. Suitable protective layers should be chosen for the purpose of excellent thermochromic properties and enhanced environmental stabilities. Hydrophobic surfaces are regarded to be advantageous to protect $VO_2$ from deterioration; nevertheless, $VO_2$ films without additional treatment exhibits a hydrophilic surface. Organic agents based on silanes have been utilized to produce hydrophobicity for $VO_2$ coatings;[35,36] however, uniformity and stability of used organic agents is lack of consideration.

For commercial thermochromic coatings, as outdoor operation conditions are not feasible due to the long time required (at least 10 years), acceleration tests with rigorous standards exist to describe pass-fail qualification tests, which is much tougher than the real operational environment. In previous work, stability tests were carried out by putting samples under specific ambient conditions and periodically measuring their thermochromic performances.[33, 37-39] Since the experimental conditions show significant influences on the deteriorating behavior of $VO_2$ films, results obtained by different studies cannot be compared simultaneously. Meanwhile, there is none of quantitative analysis of the results in the acceleration tests to the real environment stability. Based on the experimental results, we cannot calculate the corresponding service life in real operational conditions.

In this work, $HfO_2$ layers have been incorporated on $VO_2$ films for encapsulation structures based on optical design and selection. On the one hand, the proposed structure is expected to exhibit enhanced thermochromic performances due to the suitable refractive index of $HfO_2$; on the other hand, natural hydrophobicity, relatively low water vapor transmission rate and excellent mechanical properties make $HfO_2$ an excellent protective layer for the underlying $VO_2$ films. Meanwhile, four sample models have been investigated in different operational conditions to demonstrate the effect of different interfaces on the deterioration process of $VO_2$ films. Quantitative evaluation of service life has been conducted by using Hallberg-Peck acceleration model. Three types of stability tests have been carried out to imitate different operational environments of $VO_2$ films for various applications.

**RESULTS AND DISCUSSION**

*Structure Designing*

To design stable structures, we need to detect the main contributor to the instability of the existing $VO_2$ to moisture. On the surface of oxides, the coordinatively unsaturated surface oxygen plays the role as Lewis bases, which can combine with water hydrogen cations to form hydrogen bonds. Similarly, the coordinatively unsaturated metal cations can act as Lewis acids and combine with the water oxygen anions. These Lewis pairs result in the interaction between oxide surface and water molecules, and thus to hydrophilicity of the oxide surface.[40] As shown in **Figure 1A**, the interaction between $VO_2$ surface and water molecules is facile, which can be attributed to the large number of strong coordinate bonds between surface oxygen and water hydrogen, as well as the bonds between surface metal cations and water oxygen. This feature results in the hydrophilicity of $VO_2$ surface and strong wetting by water (**Figure 1B**). As for the $HfO_2$ layer, the low electronegativity of hafnium leads to the poor activity as Lewis acid. Therefore, the $HfO_2$ surface only can form weak hydrogen bonds with water hydrogen, which can be revealed by much less and directional charge density as shown in **Figure 1A**. As a result, natural hydrophobicity of $HfO_2$ layers can be observed by the water droplet contact angle measurement (**Figure 1B** and **S1**).

In addition to wettability, water vapor transmission rate (WVTR) is an important index for moisture permeation barriers. Amorphous $HfO_2$ shows relatively low water vapor transmission rate (WVTR), which makes it an excellent candidate as moisture barrier layers for $VO_2$ films.[41-42] Meanwhile, $HfO_2$ is a material with high melting point, high dielectric constant and excellent mechanical properties. In this case, the suitable refractive index of $HfO_2$ layer can meet the requirement for the antireflection effect of $VO_2$ thermochromic coatings (A Comparison of different coating materials is summarized in **Table S1**). Due to these features, the proposed $VO_2/HfO_2$ structure is expected to enhance the stability of $VO_2$ films for extended service life with stable phase transition and enhanced thermochromic performances. To explore the duration mechanism of $VO_2$ films, it is indispensable to investigate the effect of different surface and cross-section on the whole stability, which is always ignored in previous studies. In this work, as shown in **Figure 1C**, four types of $VO_2$-based films with diverse protective strategies have been fabricated. By deliberately exposing different surface and cross-section of $VO_2$ films, a detail deterioration process of $VO_2$ films is expected to be observed.

*Structure Characterizations and Performance*

Based on the structure designing above, we fabricated a novel $VO_2$/$HfO_2$ encapsulation structure by magnetron sputtering, where clear interfaces can be observed between different layers and the substrate (**Figure 2A**). It is well known that thermochromic properties of $VO_2$ films are highly dependent on relative crystallinity. Therefore, structural characterizations have been carried out of as-prepared samples. The XRD patterns of bare $VO_2$ films shows a series of peaks, which are consistent with the monoclinic phase of $VO_2$ (JCPDS: 72-0514) without any impurity peak. After the introduction of $HfO_2$ layers, the sample still exhibits a basic $VO_2$ phase and no peak belonging to crystalline $HfO_2$ can be detected, which indicates an amorphous phase of $HfO_2$ (**Figure 2B**). Raman spectroscopy and X-ray photoelectron spectroscopy also have been used for further structural characterization, where well-crystallized and stoichiometric $VO_2$ (M) films have been demonstrated to achieve desirable thermochromic performance (**Figure 2C & S2**).

Transmittance spectra of the optimal $VO_2$/$HfO_2$ structures have been shown in **Figure 2D,** where the performance of a single layer $VO_2$ film also has been exhibited for comparison. The experimental results are in accordance with the simulated results, where an optimal thickness of $HfO_2$ layer ~ 80 nm is preferred in both issues (**Figure S3&S4 & Table S2**). As for single layer $VO_2$ film, relatively low luminous transmittance (less than 40%) both in the semiconducting and metallic state can be observed due to the absorption in the short wavelength region. Meanwhile, corresponding solar modulation ability is not satisfactory with $\Delta T_{sol} = 8.8\%$, which is insufficient for energy-saving effect. After introducing ~80 nm $HfO_2$ antireflection layer, the $VO_2$/$HfO_2$ bilayer structures exhibits significantly enhanced luminous transmittance in the visible region with $T_{lum,lt} = 55.8\%$ and $T_{lum,ht} = 43.6\%$. More importantly, it is worthy to note that the proposed structure shows excellent solar modulation ability ($\Delta T_{sol} = 15.9\%$), where an improvement of 80.1% can be achieved compared to the bare sample. Enhanced luminous transmittances of the structure can also be directly observed by naked eyes, where the single layer $VO_2$ film shows dark yellow tone and the $VO_2$/$HfO_2$ bilayer structure exhibits much lighter color with acceptable transmittance (**Figure 2E**).

*Study of Environmental Stability*

To examine the environmental stability of $VO_2$ films, samples with four structures were placed in a harsh environment with high temperatures (60°C) and relative humidity (90%) for an accelerated test (**Figure 3A**). The stability of $VO_2$ films was evaluated by measuring the transmittances periodically and calculating relative solar modulation abilities and optical contrast at 2500 nm, which are important indexes to reveal phase transition feature of $VO_2$ (**Figure 3B & 3C**). As for the bare $VO_2$ film (V-1), the tough test condition makes films a noticeable deterioration. The bare sample V-1 shows a rapid loss of its thermochromic performance with weakened phase transition feature, which is totally invalid after 15 days. By encapsulating four cross-section of $VO_2$ film by silicone weatherproof sealant, the sample V-2 exhibits enhanced stability compared with the sample V-1. However, the sample V-2 still gradually lost its phase transition property with increased exposure time owing to the exposed surface. As for the sample VH-1, with the protection from the HfO$_2$ layer, a stable thermochromic performance can be maintained for about ~20 days. However, a shaper deterioration can be observed after 30 days, which can be attributed to the deterioration from the cross-section and then lead to total failure of the thermochromic performance. This phenomenon means that the cross-section plays an critical role in the deteriorating process of VO$_2$ films, as important as the surface and even more, in spite of its relatively small area (1:200000 in this case, **Table S3**). Surprisingly, the sample VH-2 with full encapsulation by HfO$_2$ layer shows robust stability during the accelerated test, while there is an indiscernible change in corresponding transmittance spectra, even after a long aging time of 100 days. Detailed transmittance spectra have been depicted in **Figure S5**.

For quantitative evaluation of the service life, there must be a reliable relationship between the testing conditions and the operating conditions. The most acceptable acceleration model is proposed by Hallberg and Peck,[43] which is developed from the Arrhenius's equation to combine the effects of temperature and humidity. The Hallberg-Peck model is given by

$$AF = \exp\left[\frac{E_a}{k} * \left(\frac{1}{T_{use}} - \frac{1}{T_{test}}\right)\right] * \left(\frac{RH_{test}}{RH_{use}}\right)^n$$

Where $AF$ is the acceleration factor, $E_a$ is the activation energy (eV), $k$ is Boltzmann's constant (8.617×10$^{-5}$ eV/K), $T_{use}$ and $T_{test}$ is the operating temperature (K) and the acceleration test temperature (K), respectively; $RH_{use}$ and $RH_{test}$ is the acceleration test humidity and the operating humidity, respectively. In this case, $T_{test} = 60°C = 333K$ with a relative humidity $RH_{test} = 90\%$. Based on field data, the nominal operating temperature and humidity are $T_{use} = 25°C = 298K$ and $RH_{use} = 60\%$; the activation energy for VO$_2$ is taken to be 0.7 eV [44]. According to the model, the recommended values for *n* is 3. The acceleration factor for the aging test was calculated by substituting the given parameters in the Hallberg-Peck model, yielding $AF$ = 59.24. In our accelerated aging test, VO$_2$ films fully encapsulated by HfO$_2$ can resist the testing environment with 60°C and 90% relative humidity after exposure time $T$ of 100 days. Therefore, the guaranteed life of the proposed VO$_2$/HfO$_2$ structure is equal to $AF \times T$ = 59.24 × 100 days = 5924 days ≈ 16 years, which can satisfy the demand for practical applications.

We also designed an extremely tough test by placing samples in boiling water to examine their resistance to moisture (**Figure S6**). Photography of samples after different treated time in boing water and relative change in solar modulation abilities have been shown in **Figure 4A & 4B**, while transmittance spectra have been illustrated in **Figure S7**. The bare film V-1 shows a fast failure of its thermochromic performance during the boiling process and being invalid only after 8 hours. The sample VH-1 with exposed cross-section exhibits largely enhanced stability compared to bare VO$_2$. Nevertheless, the sample VH-2 fully encapsulated by HfO$_2$ shows the best stability among these samples. Even being placed in boiling water of 24 hours, about 56% solar modulation ability (8.9% to 15.9%) can be maintained. Largely enhanced stability of the encapsulated VO$_2$/HfO$_2$ structure to moisture can be summarized to following reasons: On the one hand, the HfO$_2$ layer is water resisting and can prevent the bottom VO$_2$ films from moisture and following rapid deterioration. On the other hand, the proposed fully encapsulated structures provide comprehensive protection and largely enhanced service life of $VO_2$ films. This also revealed that the protected cross-section is helpful for enhanced stability of $VO_2$ films.

*Study of Thermal Stability*

Thermal stability is another parameter that should be taken into consideration for practical applications of $VO_2$ films, because $VO_2$ can be oxidized by oxygen, especially at high temperatures. Thermal stability of samples were examined by annealing at high temperatures range from 300 to 400°C. The characteristic (011) peak of $VO_2$ (M) was investigated to study the thermal stability of samples at different temperatures (**Figure 4C**), while corresponding solar modulation abilities have been depicted in **Figure 4D**. With increased treatment temperatures, the bare film V-1 shows obviously deteriorated thermochromic performances since 300°C, and totally loses its phase transition property at 375°C. According to corresponding XRD patterns, the intensity of the $VO_2$ (011) peak become weaker than the pristine sample with increased annealing temperatures. After heating treatment at 375°C, the sample transformed into the $V_2O_5$ phase, which does not possess thermochromic performance near room temperature (**Figure S8**). For the sample VH-1, corresponding thermochromic performance can be maintained until 350°C without any deterioration but still totally become invalid at 400°C (**Figure 5D**). The film is stable with pure monoclinic $VO_2$ phase up to 350°C and then transformed into the $V_2O_5$ phase with increased temperatures, which is in accordance with the optical properties. It is worthy to note that for the sample treated at 375°C and above, the peak around ~28° is not the (011) peak of $VO_2$, but the (-111) peak of the monoclinic $HfO_2$, which is crystallized with increased temperature. On contrast, the sample VH-2 shows enhanced thermal stability up to 375°C with stable thermochromic performance (**Figure 5D** and **S9**) and pure monoclinic phase of $VO_2$. The invalid thermochromic performances after thermal treatment at 400°C can be attributed to the grain boundaries of crystalline $HfO_2$ layers (**Figure S10**), which can serve as gas-diffusion pathways and lead to the failure of underlying $VO_2$ films. Different thermal stability depicted by above samples indicates that the cross-section plays an important role in the deterioration process of $VO_2$ films, and that the encapsulated $HfO_2$ layer can sufficiently protect the $VO_2$ films from oxidation. $VO_2$ films with enhanced thermal stability can serve as special applications in high temperature environments, such as thermal sensors, laser protections, etc.

*Mechanism Discussion*

Based on above experimental results, an interesting phenomenon can be observed that the sample with protected surface but exposed cross-section shows a fast deteriorating process and the cross-section plays a critical role in the deterioration process of $VO_2$ films. By fully encapsulated by the $HfO_2$ layer, the $VO_2/HfO_2$ structure can obtain a largely enhanced service life in different operational conditions. The microstructure of deposited films is considered to be an important factor that influences the deteriorating process of $VO_2$ films. **Figure 5** reveals schematic representation of dependence of film structure on substrate temperature and argon pressure during the magnetron sputtering process. [45] One axis is the argon pressure during the sputtering process, another axis is the $T/T_m$, where $T$ is the substrate temperature and $T_m$ is the film materials melting point ($K$). In this case, $T$ of substrate temperature during sputtering process of $VO_2$ films is at ~450°C, while the melting point $T_m$ of $VO_2$ is 1640°C. The deposition pressure is about 6 mTorr. Therefore, the microstructure of the deposited $VO_2$ films can be located, as shown in the right circle with an enlarged vision. In this zone, the film consists of packed fibrous grains with a relatively smooth

"fine-domed" surface. In the cross-section of the films, the grains were surrounded by several micro holes and cracks, which can act as "channels" to let the moisture and oxygen pass through and leads to deterioration of the films. Therefore, for an enhanced service life, the cross-section of the $VO_2$ films cannot be ignored and should be protected as well as the film surface. By deliberately encapsulated the cross-section and surface of fabricated $VO_2$ films by $HfO_2$ protective layer, the proposed structure shows greatly enhanced service life for practical applications.

**EXPERIMENTAL PROCEDURES**

The samples were grown on 10×10 mm² quartz glass substrates by magnetron sputtering system. The $VO_2$ films were deposited via sputtering using a $V_2O_3$ ceramic target (purity 99.9%, 4-inch diameter) under the condition of 100 W DC power with Ar and $O_2$ flows of 39 and 1 sccm, respectively. The deposition of $HfO_2$ films was carried out using an $HfO_2$ ceramic target (purity 99.9%, 4-inch diameter) of 150 W RF power with 40 sccm argon flow (purity 99.99%). The pressure of the deposition chamber was maintained at 6.00 mTorr, and the substrate temperature was kept at 450°C for $VO_2$ and at room temperature for $HfO_2$. The substrate temperatures were systematically calibrated by a surface thermometer before the deposition. In this work, four types of $VO_2$-based samples have been fabricated: bare $VO_2$ films directly deposited on quartz substrate without any protection (denoted as V-1); $VO_2$ films with encapsulated sides by ~2 mm silicone weatherproof sealant, while the surface was direct exposure to air (denoted as V-2); The top surface and three cross-sections of $VO_2$ film were protected by $HfO_2$ layer, while one cross-section was deliberately uncovered to investigated its effect on the deterioration process of $VO_2$ films (denoted as VH-1); and $VO_2$ films fully encapsulated by $HfO_2$ protective layer (denoted as VH-2).

As-prepared samples were subjected to structural characterization and property studies. Thin film X-ray diffraction (XRD) analysis was conducted on a Rigaku Ultima IV diffractometer with Cu Kα radiation ($\lambda$=1.5418 Å) using 2θ scanning model. The morphology and thickness of the films were measured by a scanning electron microscope (SEM, Hitachi SU8220) and an atomic force microscope (AFM, SII Nano Technology Ltd, Nanonavi P). X-ray photoemission spectroscopy (XPS) analysis was performed on Thermo Fisher Scientific ESCAlab250 to investigate the valance state of elements in the films. Transmittances spectra of the bilayer structures in the wavelength range of 350 to 2600 nm were measured using a UV-vis spectrophotometer (Hitachi U-3100). Raman analysis was carried out using the HR Evolution Raman spectrometer.

$VO_2$ (M) thin films have been widely studied to be utilized as smart windows. For the investigation of the optical properties, vis-near-infrared transmittances spectra of $VO_2/HfO_2$ bilayer films and single layer $VO_2$ films deposited at different temperatures were measured. The integral luminous transmittances (380-780 nm) and solar transmittances (350-2600 nm) of the films were obtained by the following equations:

$T_{lum} = \int \Phi_{lum}(\lambda) T(\lambda) d\lambda / \int \Phi_{lum}(\lambda) d\lambda$ (1)

$T_{sol} = \int \Phi_{sol}(\lambda) T(\lambda) d\lambda / \int \Phi_{sol}(\lambda) d\lambda$ (2)

where $T(\lambda)$ represents the transmittance at wavelength $\lambda$; $\Phi_{lum}$ is the standard efficiency function for photopic vision; and $\Phi_{sol}$ is the solar irradiance spectrum for an air mass of 1.5, which corresponds to the sun standing 37° above the horizon. While the solar modulation ability ($\Delta T_{sol}$) of the films was calculated by $\Delta T_{sol} = T_{sol,lt} - T_{sol,ht}$, where $lt$ and $ht$ represent 25°C and 90°C, respectively.

To examine the environmental stability of multilayer films, different samples were placed in a constant-temperature humidity chamber with the accelerated experimental conditions of $T_{test} = 60°C$ and relative humidity $RH_{test} = 90\%$, and optical transmittance measurements were carried out to estimate the optical durability. An extreme environmental stability test also has been conducted by bathing samples in boiling water and periodically measuring optical transmittances. To investigate the thermal stability, many samples have been prepared and heated in air using a muffle furnace from room temperature to 300 - 400°C with a heating rate of 50°C/min and maintained for 20 min. After heating, the samples were naturally cooled down to room temperature for optical performances measurements.

## SUPPLEMENTAL INFORMATION
Supplemental Information includes ten figures and three tables.

## AUTHOR CONTRIBUTIONS
T.C., X.C., and H.L designed the experiments and analyzed the data. T.C., N.L., Y.Z. and S.L. performed the experiments and some characterizations. T.C. performed optical simulations under the supervision of P.J. T.C. and X.C. drafted the paper. All authors discussed the results and commented on the manuscript.

## ACKNOWLEDGMENTS

This study was financially supported by the National Natural Science Foundation of China (NSFC, No.51572284), the Youth Innovation Promotion Association, Chinese Academy of Sciences (No.2018288), the Science Foundation for Youth Scholar of State Key Laboratory of High Performance Ceramics and Superfine Microstructures (No.SKL201703), the Shanghai Pujiang Program (No. 18PJD051) and the Key Research and Development Plan of Anhui Province (1804a09020061).

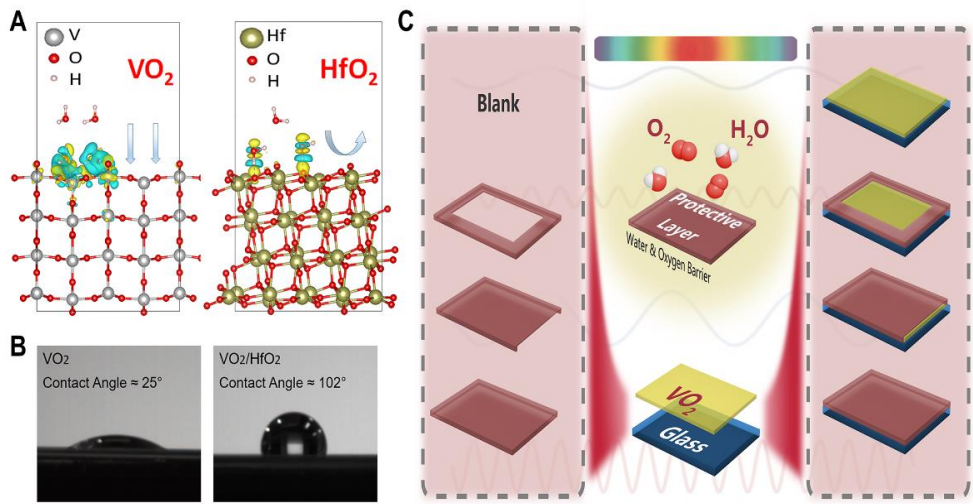

**Figure 1. Water molecule adsorption on different oxide surface and structure designing.**

(A) Possible interaction between water molecules and vanadium dioxide surface and hafnium dioxide surface, respectively. V, Hf, O and H atoms are represented by balls in grey, blue, red and white, respectively.

(B) Water contact angle measurement of different oxide films. As for the $VO_2$ film, the static water contact angle is about 25.0°, which means a hydrophilic surface. The introduction of the $HfO_2$ layer makes the static water contact angle of the films changed abruptly from 25.0° (hydrophilicity) to 102.0° (hydrophobicity).

(C) Schematic illustrations of four types of samples.

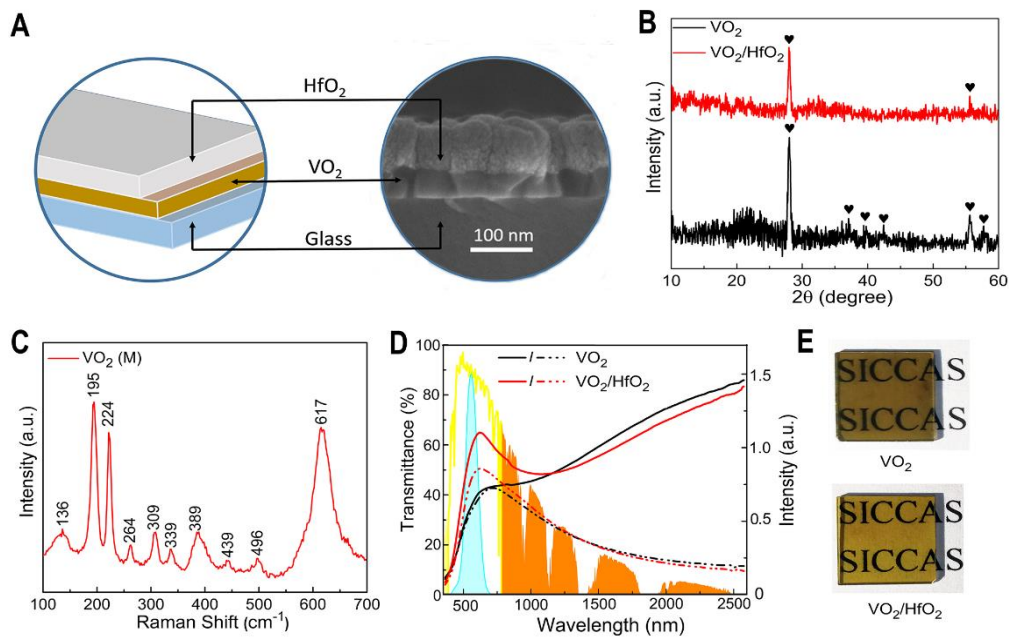

**Figure 2. Characterizations and Performance of $VO_2/HfO_2$ films.**

(A) Schematic that shows the device architecture and corresponding cross-section SEM micrograph of the $VO_2/HfO_2$ bilayer film.

(B) X-ray diffraction patterns of $VO_2$ single layer film and $VO_2/HfO_2$ bilayer film.

(C) Raman spectra of as-prepared $VO_2$ film without additional $HfO_2$ layer.

(D) Transmittance spectra (350–2600 nm) at 25°C (solid lines) and 90 °C (dashed lines) of $VO_2$ film and $VO_2/HfO_2$ bilayer film.

(E) Photography of $VO_2$ film and $VO_2/HfO_2$ bilayer film on quartz glass substrates.

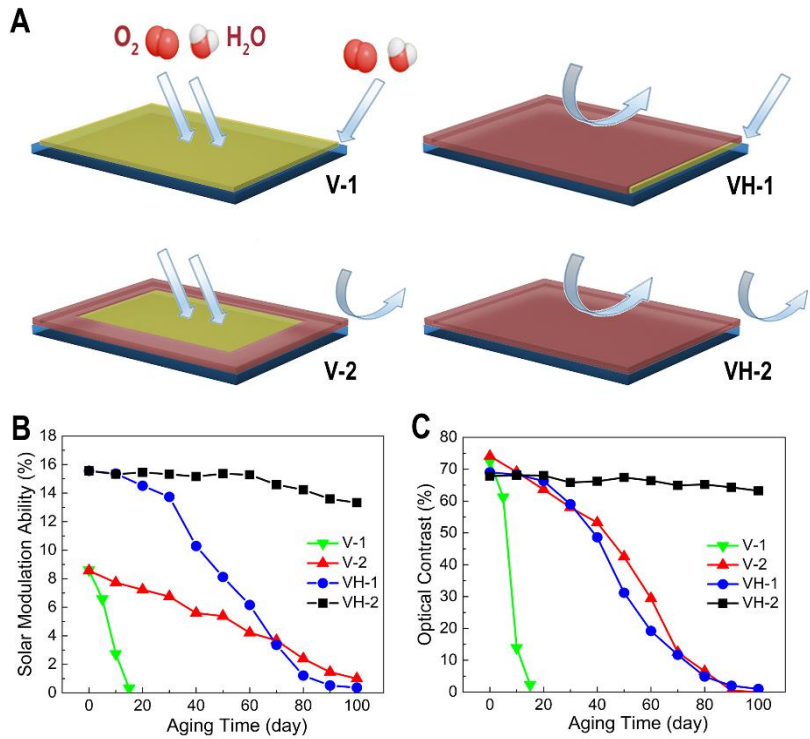

**Figure 3. Environmental stability test of different samples.**

(A) Schematic illustration of the possible effect of oxygen and moisture on different samples.

(B) Solar modulation abilities versus increasing the aging time for different samples.

(C) Optical contrast at 2500 nm in the infrared region versus increasing aging time for different samples.

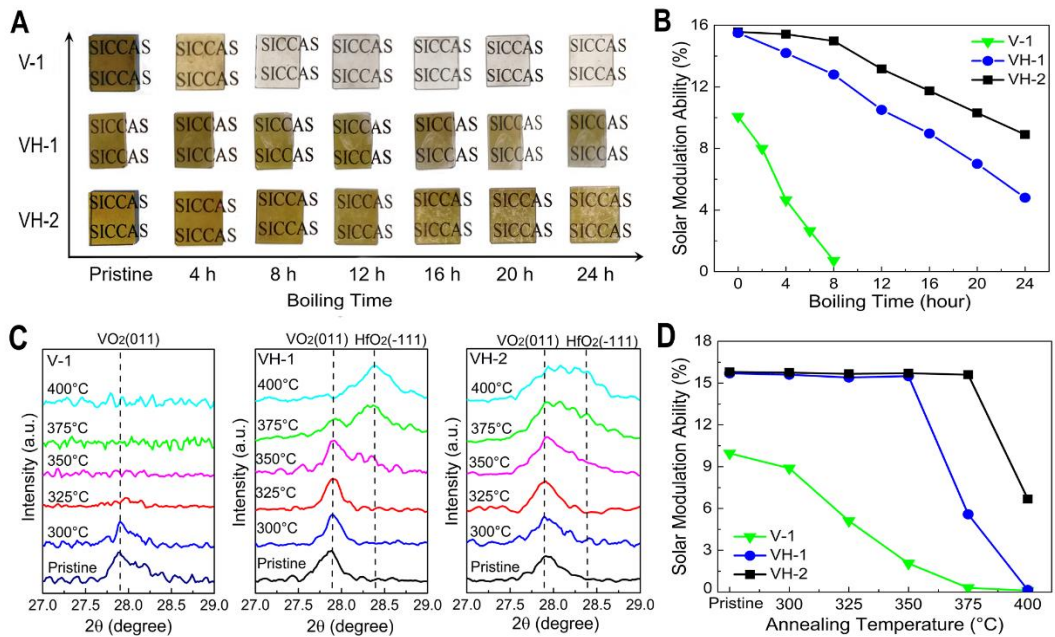

**Figure 4. Boing water test and thermal stability test of samples.**

(A) Photography of samples after different treatment time in boing water.

(B) Solar modulation abilities of different samples versus increasing treatment time in boing water.

(C) XRD patterns of samples after thermal treatment at different annealing temperatures.

(D) Solar modulation abilities of different samples versus increasing annealing temperatures.

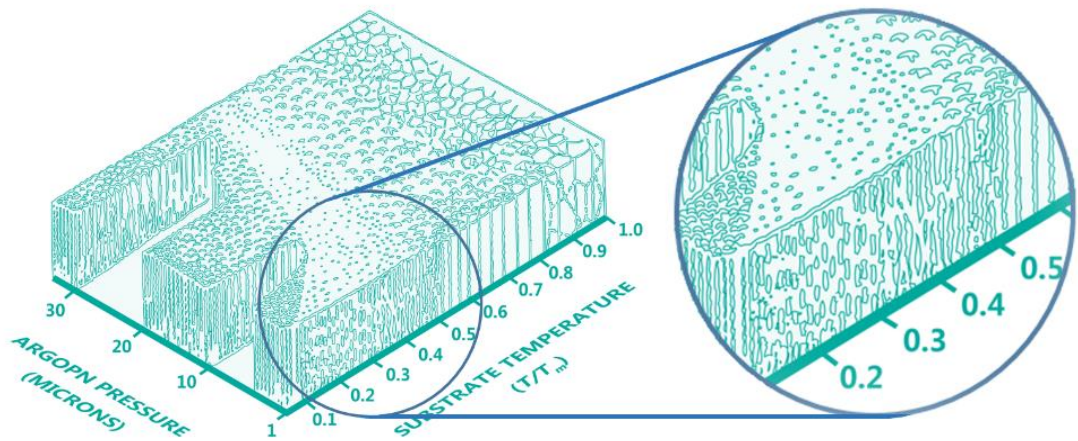

**Figure 5. Schematic representation of dependence of coating structure on substrate temperature and argon pressure.** The microstructure of deposited films is depended on two parameters: One is the argon pressure during the sputtering process, another is the $T/T_m$, where $T$ is the substrate temperature and $T_m$ is the film materials melting point (K).

# Supplemental Information

# Towards Commercializing Vanadium Dioxide Films: Investigation of the Impact of Different Interface on the Deterioration Process for Largely Extended Service Life

Tianci Chang,[1,2] Xun Cao,[1,*] Ning Li,[3] Shiwei Long,[1,2] Ying Zhu,[1,2] Jian Huang,[1] Hongjie Luo,[4] Ping Jin[1,5,*]

[1] State Key Laboratory of High Performance Ceramics and Superfine Microstructure, Shanghai Institute of Ceramics, Chinese Academy of Sciences, Shanghai 200050, China

[2] University of Chinese Academy of Sciences, Beijing 100049, China

[3] Department of Materials Science and Engineering, College of Science, China University of Petroleum Beijing, Beijing 102249, China

[4] School of Materials Science and Engineering, Shanghai University, Shanghai 200444, China

[5] National Institute of Advanced Industrial Science and Technology (AIST), Moriyama, Nagoya 463-8560, Japan

[*] Correspondence: cxun@mail.sic.ac.cn (X.C.), p-jin@mail.sic.ac.cn (P.J.)

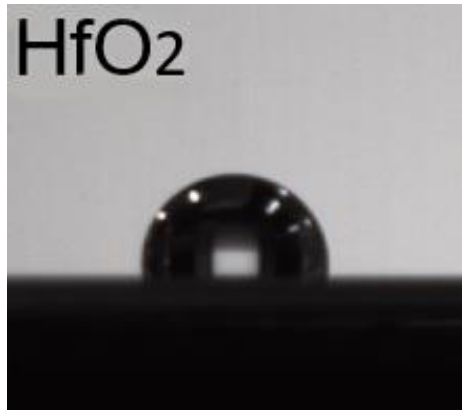

**Figure S1.** Water contact angle measurement of single layer HfO$_2$ layer without VO$_2$ films. As for the single layer HfO$_2$, the measured water contact angle was about 106°, which indicates a hydrophobic surface for water resisting function.

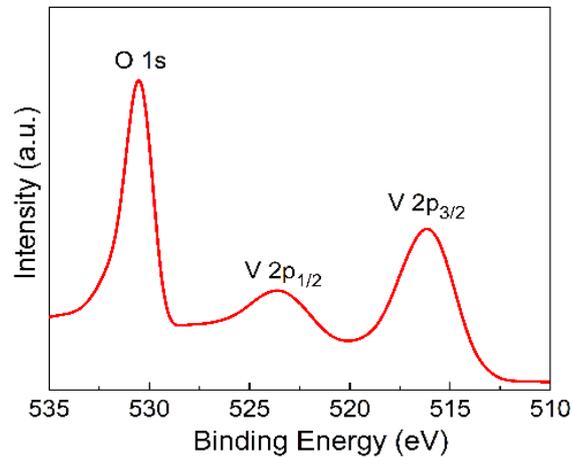

**Figure S2. XPS Characterization of VO$_2$ film.** Figure S3 shows V 2p and O 1s photoelectron spectra of the VO$_2$ film. The position of the V 2p$_{3/2}$ core-level peak in the spectra is ~516.1 eV while the peak of V 2p$_{1/2}$ is at 523.5 eV, which can be attributed to the V$^{4+}$ in VO$_2$. The binding energy of O 1s appears at 530.5 eV. The results of XPS suggest that the valence of vanadium is +4 and confirms that the obtained vanadium oxide films is stoichiometric and consists of pure VO$_2$ phase.

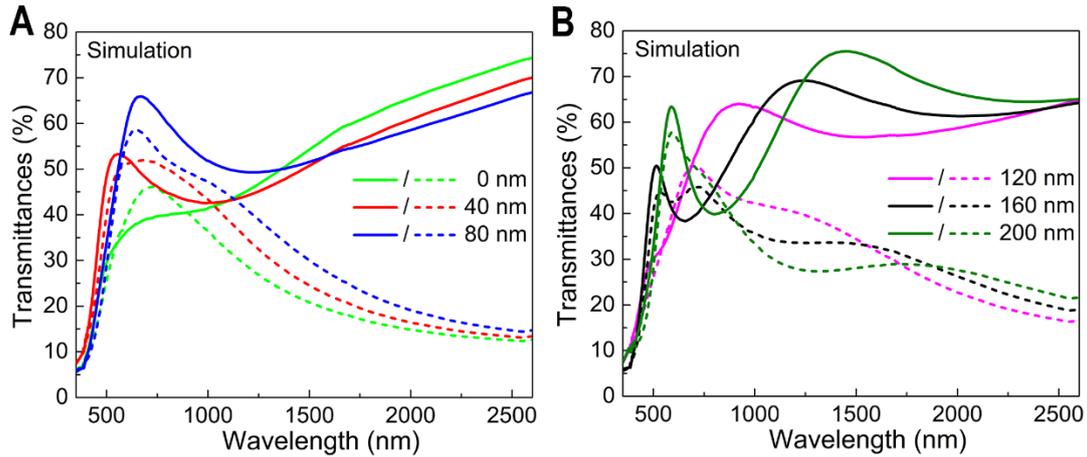

**Figure S3. Optical simulation for optimal structures.** It is important to know the suitable value of the thickness of $HfO_2$ layer for $VO_2$ towards the goal of obtaining the best optical performance of the proposed structure. For the purpose, an optical calculation and optimization was conducted on the Essential Macleod software. The optical constants of the films were measured using ellipsometry on films deposited on glass substrates with fixed thickness of $VO_2$ (50 nm) and various thickness of $HfO_2$ (0-200 nm). The thickness of $HfO_2$ layer shows remarkable influence on the optical performances of $VO_2$ films, and the luminous transmittance can be largely enhanced when the thickness of $HfO_2$ at 80 nm. With continuously increased thickness of the $HfO_2$ layer, the transmittances spectra of the structure demonstrated a wave-like feature owing to the interference effect of the structure.

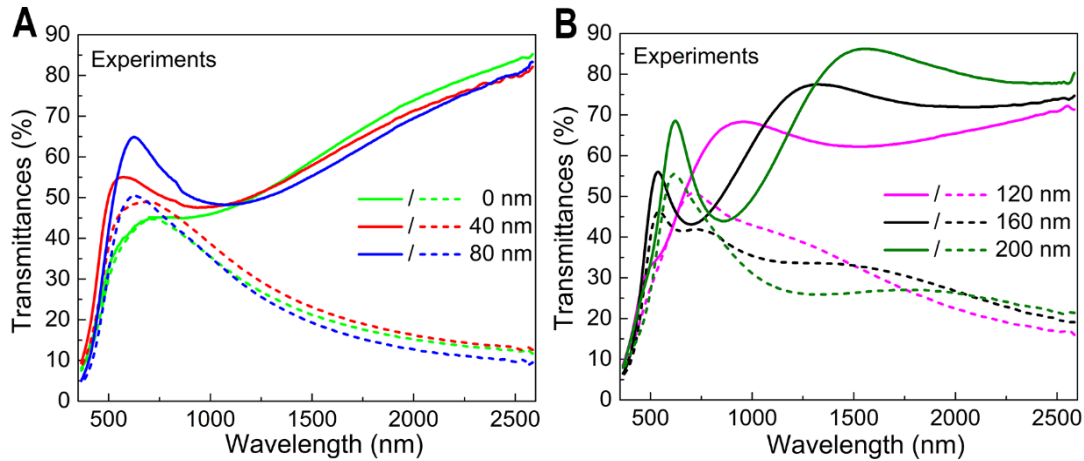

**Figure S4. Transmittances spectra of $VO_2/HfO_2$ structures.** The thickness of the bottom $VO_2$ layer is fixed at ~50 nm for optimized optical performance. The top $HfO_2$ layer is deposited with various thickness (0-200 nm) to investigate the influence on the optical performance of whole structures. The change of optical transmittance is in accordance with the simulated results. When the thickness of $HfO_2$ layer increased from 0 nm to 80 nm, the $VO_2/HfO_2$ structure showed largely enhanced luminous transmittances in the visible region due to the antireflection effect of $HfO_2$ layer to the $VO_2$ layer. With continuously increased thickness of the $HfO_2$ layer, the transmittances spectra of the structure demonstrated a wave-like feature owing to the interference effect of the structure for both semiconducting and metallic $VO_2$.

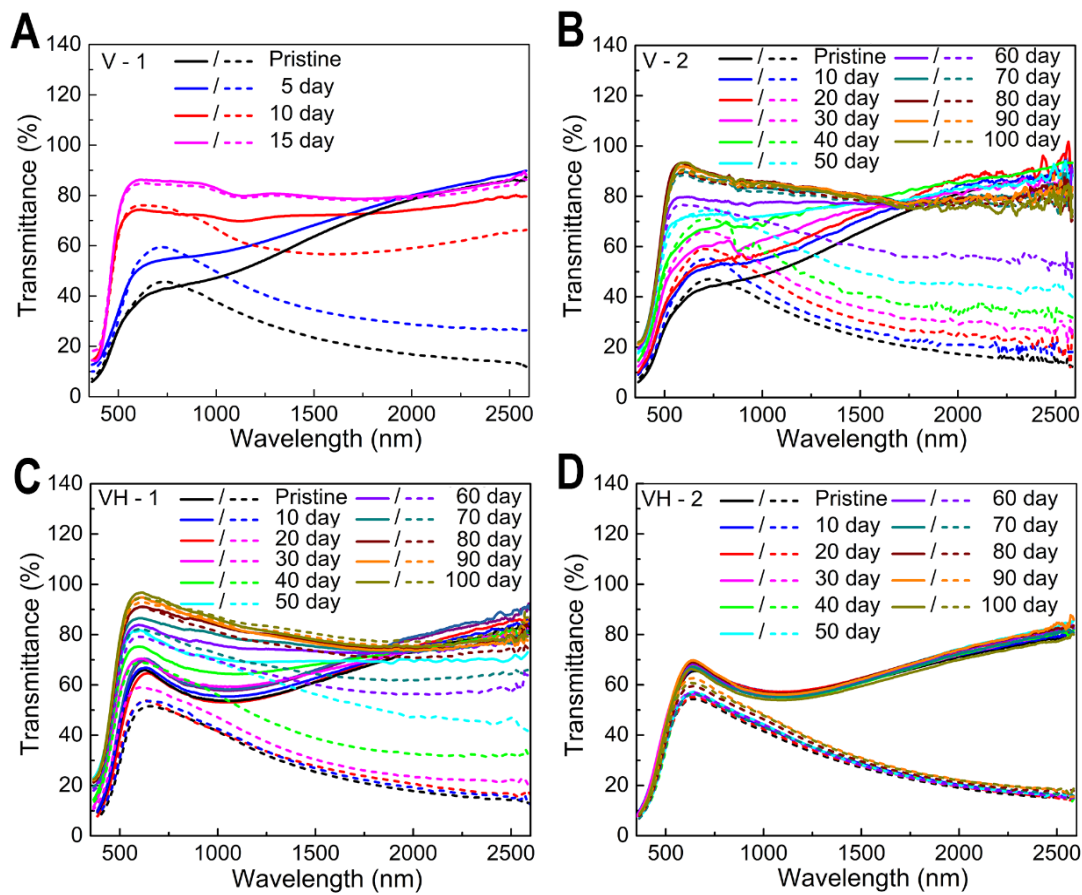

**Figure S5. (A)-(D) Transmittance spectra at 25°C (solid lines) and 90°C (dashed lines) of samples after different duration time (0-100 day) for V-1, V-2, VH-1 and VH-2, respectively.** The test was conducted under harsh environment with high temperature (60°C) and relative humidity (90%). The unprotected sample V-1 shows a rapid loss of its thermochromic performance. The sample V-2 exhibits enhanced stability compared with the bare film, but still gradually lost its phase transition property with increased exposure time owing to the exposed surface. The sample VH-1 shows a stable thermochromic performance, which can be maintained for about ~20 days. However, a shaper deterioration of thermochromic performance can be observed after 30 days than the sample V-2. The sample VH-2 with full encapsulation by $HfO_2$ layer shows robust stability during the accelerated aging test, even after a long aging time of 100 days.

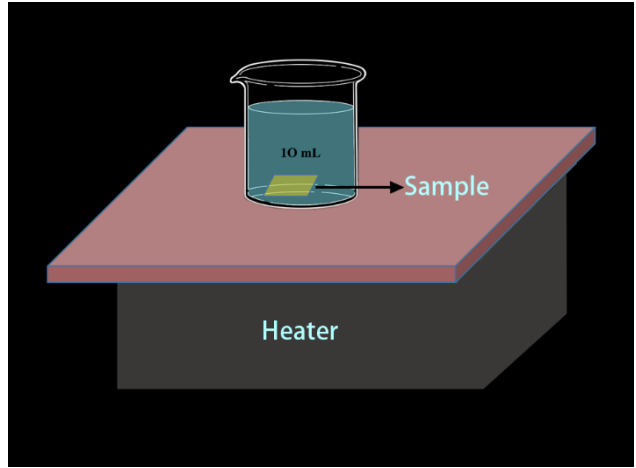

**Figure S6. Schematic illustration of the boiling water test.** We set an extremely harsh environment to test the resistance of the samples to moistures. The samples were put in a beaker with deionized water and a heater was used to heat the water to boiling point. The temperature of the heater was set at 200°C to keep the water boiling. Optical transmittances of the samples were measured periodically to evaluate the duration process.

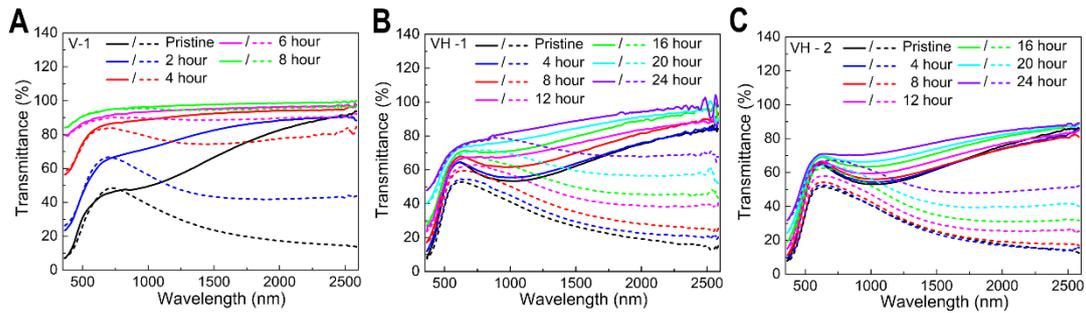

**Figure S7. Transmittance spectra of samples after different treatment time in boing water.** The bare V-1 film cannot resist the boing water and shows an abrupt deteriorated process only in 8 hours. The sample VH-1 exhibits better stability than the single layer film due to the protect effect of $HfO_2$. However, owing to the fully encapsulated surface and cross-section, the sample VH-2 shows the best resistance to boing water and still exhibits ~9% solar modulation ability even after 24 hours treatment in boiling water.

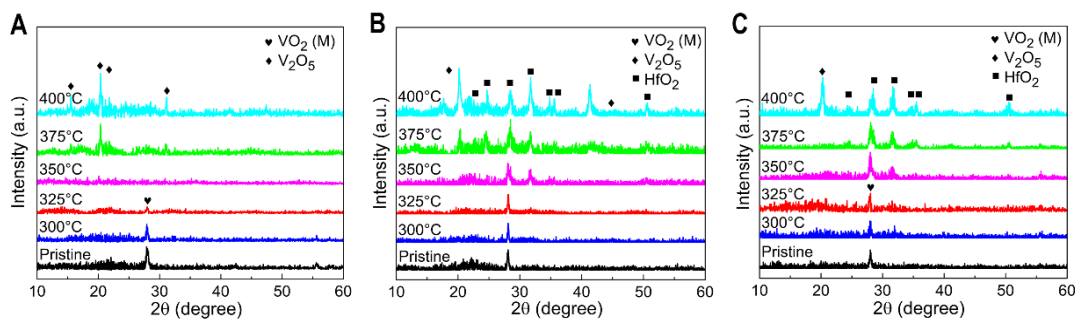

**Figure S8. XRD patterns of VO$_2$/HfO$_2$ films after thermal treatment at different temperatures.** According to corresponding XRD patterns, as for the sample V-1, the intensity of the VO$_2$ (011) peak become weaker than the pristine sample with increased heating temperatures. After heating treatment at 375°C, the sample transformed into the V$_2$O$_5$ phase, which not possesses thermochromic performance near room temperature. For the sample with HfO$_2$ protective layer but exposed cross-section (VH-1), the film is stable with pure monoclinic VO$_2$ phase up to 350°C and then transformed into the V$_2$O$_5$ phase with increased temperatures, which is in accordance with the optical properties. On contrast, the VO$_2$ film with encapsulated protective layer (VH-2) shows enhanced thermal stability up to 375°C with stable thermochromic performance and pure monoclinic phase of VO$_2$.

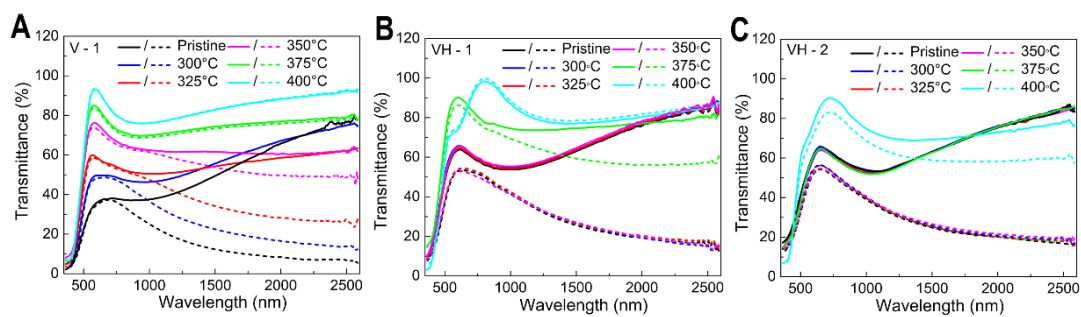

**Figure S9. (A)-(C) Transmittance spectra at 25°C (solid lines) and 90°C (dashed lines) of samples after thermal treatment at different temperatures (300-400°C) for V-1, VH-1 and VH-2, respectively.** With increased treatment temperatures, the bare V-1 film shows obviously deteriorated thermochromic performances since 300°C, and totally lose its phase transition property at 375°C. For the sample with HfO$_2$ protective layer but exposed cross-section (VH-1), corresponding thermochromic performance can be maintained until 350°C without any deterioration but still totally being invalid at 400°C. On contrast, the VO$_2$ film with encapsulated protective layer (VH-2) shows enhanced thermal stability up to 375°C with stable thermochromic performance. This means that the cross-section plays an important role in the deterioration process of VO$_2$ films, and the encapsulated HfO$_2$ layer can sufficiently protect the VO$_2$ films from oxidation.

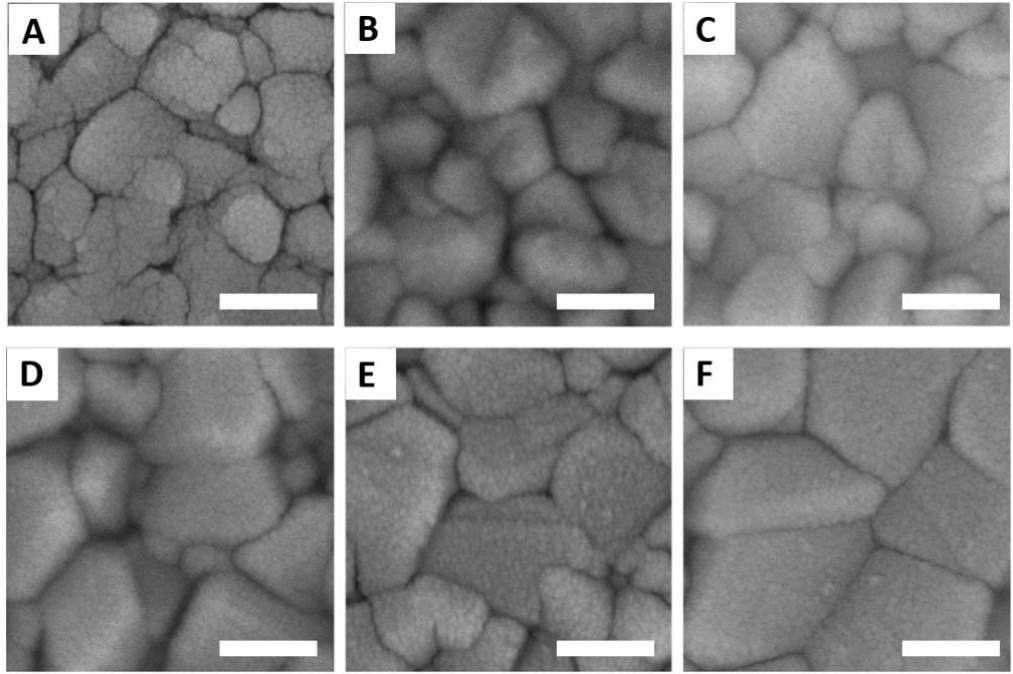

**Figure S10. SEM images of VO$_2$/HfO$_2$ films after thermal treatment at different temperatures.** The scale bar in all images is 100 nm. (A) Pristine sample, (B)-(D) samples after thermal treatment at 300°C, 325°C, 350°C, 375°C and 400°C, respectively. Due to the amorphous structure of HfO$_2$ layer, the pristine sample is consisted of condensed nanoparticles without obvious grains. With increased thermal treatment temperatures, the sample shows larger particle sizes and grain boundaries. After thermal treatment at 400°C, the grain size reached to about ~150 nm and obvious boundaries.

**Table S1.** Comparison of different coating materials.

| Coating type | Refractive index | Hardness (GPa)* | Wettability |
|---|---|---|---|
| $SiO_2$ | 1.45 [1] | 5.0 [2] | Hydrophilicity[3] |
| $TiO_2$ | 2.50 [4] | 3.5 [5] | Hydrophilicity[6] |
| $SnO_2$ | 1.60 [7] | ~5.0 [8] | Hydrophilicity[9] |
| $WO_3$ | 2.1 [10] | 4.0 [11] | Hydrophilicity[9] |
| $ZnO$ | 1.9-2.1 [12] | 4-6 [13] | Hydrophilicity[9] |
| $Al_2O_3$ | 1.5-1.6 [14] | 7-8 [15] | Hydrophilicity[16] |
| $CeO_2$ | 2.3-2.4 [17] | 11.7 [18] | Hydrophilicity[19] |
| $Si_3N_4$ | 1.97 [20] | ~12 [21] | Hydrophilicity[22] |
| $ZrO_2$ | 2.0 [23] | 6-12 [24; 25] | Hydrophilicity[25] |
| $HfO_2$ | 2.0-2.2 [26] | 8-15 [2; 27] | Hydrophobicity[28] |

* The samples in references were amorphous.

**Table S2.** Optical properties of $VO_2/HfO_2$ structures with different thickness of the $HfO_2$ layer.

| Sample* | $T_{lum,lt}(\%)$ | $T_{lum,ht}(\%)$ | $T_{sol,lt}(\%)$ | $T_{sol,ht}(\%)$ | $\Delta T_{sol}(\%)$ |
|---|---|---|---|---|---|
| $VO_2$ | 38.1 | 36.6 | 41.5 | 32.7 | 8.8 |
| $VO_2/HfO_2$ (40 nm) | 53.5 | 44.6 | 49.3 | 37.3 | 12.0 |
| $VO_2/HfO_2$ (80 nm) | 55.8 | 43.6 | 50.5 | 34.6 | 15.9 |
| $VO_2/HfO_2$ (120 nm) | 37.3 | 35.4 | 50.8 | 37.2 | 13.6 |
| $VO_2/HfO_2$ (160 nm) | 51.4 | 43.4 | 51.5 | 35.7 | 15.8 |
| $VO_2/HfO_2$ (200 nm) | 51.6 | 43.6 | 52.0 | 35.2 | 16.8 |

* The values in brackets present for the thickness of the $HfO_2$ layer of the corresponding sample.

**Table S3.** Area calculation of the samples.

| | Length | Width | Area |
|---|---|---|---|
| Surface of the $VO_2$ film | 1 cm | 1 cm | 1 $cm^2$ |
| One cross-section of the $VO_2$ film | 1 cm | ~50 nm | $5*10^{-6}$ $cm^2$ |